\documentclass[a4paper,12pt]{article}
\usepackage{epsfig}
\usepackage{amssymb}
\usepackage{amsfonts}
\usepackage{amsmath}
\usepackage{euscript}
\usepackage{verbatim}
\usepackage{latexsym}
\usepackage{graphicx}

	\usepackage{graphicx}	
	\usepackage{subfigure}
	\usepackage{wrapfig}
	\usepackage[T1]{fontenc}
	\usepackage{mathtools}
	\usepackage{float}

\usepackage{tikz}
\usetikzlibrary{shapes.geometric, arrows,patterns,snakes}
\tikzstyle{ellip} = [ellipse, minimum width=3cm, minimum height=1cm,text centered, draw=black]

\newskip\humongous \humongous=0pt plus 1000pt minus 1000pt

\newif\ifdtup

\catcode`\@=11

\@addtoreset{equation}{section}

\def\@normalsize{\@setsize\normalsize{15pt}\xiipt\@xiipt
\abovedisplayskip 14pt plus3pt minus3pt%
\belowdisplayskip \abovedisplayskip
\abovedisplayshortskip \z@ plus3pt%
\belowdisplayshortskip 7pt plus3.5pt minus0pt}

\def\small{\@setsize\small{13.6pt}\xipt\@xipt
\abovedisplayskip 13pt plus3pt minus3pt%
\belowdisplayskip \abovedisplayskip
\abovedisplayshortskip \z@ plus3pt%
\belowdisplayshortskip 7pt plus3.5pt minus0pt
\def\@listi{\parsep 4.5pt plus 2pt minus 1pt
     \itemsep \parsep
     \topsep 9pt plus 3pt minus 3pt}}

\relax

\catcode`@=12

\catcode`\@=11

\def\section{\@startsection{section}{1}{\z@}{3.5ex plus 1ex minus
   .2ex}{2.3ex plus .2ex}{\large\bf}}

\def\SymBoxes#1#2#3#4{\newdimen\un@t \un@t#3%
\raisebox{#1}{\rule{#2\un@t}{#4}\hskip-#2\un@t
\@tempdimb\un@t \advance\@tempdimb by-#4\@tempcntb#2\relax%
\@whilenum{\@tempcntb>0}\do{
\rule{#4}{\un@t}\hskip\@tempdimb \advance\@tempcntb by\m@ne}%
\hskip-#2\un@t \rule[\un@t]{#2\un@t}{#4}%
\rule[\un@t]{#4}{#4}\hskip-#4
\rule{#4}{\un@t}}\hskip-#4}                

\begin{document}

\newcommand{\beq}{\begin{equation}}
\newcommand{\eeq}{\end{equation}}
\newcommand{\bea}{\begin{eqnarray}}
\newcommand{\eea}{\end{eqnarray}}
\newcommand{\beas}{\begin{eqnarray*}}
\newcommand{\eeas}{\end{eqnarray*}}
\newcommand{\defi}{\stackrel{\rm def}{=}}
\newcommand{\non}{\nonumber}
\newcommand{\bquo}{\begin{quote}}
\newcommand{\enqu}{\end{quote}}
\renewcommand{\(}{\begin{equation}}
\renewcommand{\)}{\end{equation}}
\def \eqn#1#2{\begin{equation}#2\label{#1}\end{equation}}

\def\bA{\bar{A}}
\def\ba{\bar{a}}
\def\e{\epsilon}
\def\calM{{\mathcal M}}
\def\cL{\mathcal{L}}
\def\cW{\mathcal{W}}
\newcommand{\calL}[1] {\mathcal{L}^{(#1)}}
\newcommand{\calW}[1] {\mathcal{W}^{(#1)}}
\newcommand{\mui}[1] {\mu^{(#1)}}
\newcommand{\nui}[1] {\nu^{(#1)}}
\newcommand{\bcalL} {\overline{\cL}}
\newcommand{\bcalW} {\overline{\cW}}
\def\bmu{\overline{\mu}}
\def\bnu{\overline{\nu}}
\newcommand{\ei}[1] {\epsilon^{(#1)}}
\newcommand{\si}[1] {\sigma^{(#1)}}
\newcommand{\bei}[1] {\overline{\epsilon}^{(#1)}}
\newcommand{\bsi}[1] {\overline{\sigma}^{(#1)}}
\def\pp{\partial_{\phi}}
\def\pt{\partial_t}
\def\rme{\mathrm{e}}
\def\Tr{ \hbox{\rm Tr}}
\def\H{ \hbox{\rm H}}
\def\HE{ \hbox{$\rm H^{even}$}}
\def\HO{ \hbox{$\rm H^{odd}$}}
\def\K{ \hbox{\rm K}}
\def\Im{ \hbox{\rm Im}}
\def\Ker{ \hbox{\rm Ker}}
\def\const{\hbox {\rm const.}}
\def\o{\over}
\def\im{\hbox{\rm Im}}
\def\re{\hbox{\rm Re}}
\def\bra{\langle}
\def\ket{\rangle}
\def\Arg{\hbox {\rm Arg}}
\def\Re{\hbox {\rm Re}}
\def\Im{\hbox {\rm Im}}
\def\exo{\hbox {\rm exp}}
\def\diag{\hbox{\rm diag}}
\def\longvert{{\rule[-2mm]{0.1mm}{7mm}}\,}
\def\a{\alpha}
\def\dag{{}^{\dagger}}
\def\tq{{\widetilde q}}
\def\p{{}^{\prime}}
\def\W{W}
\def\N{{\cal N}}
\def\calB{\mathcal{B}}
\def\hsp{,\hspace{.7cm}}

\def\br{\nonumber\\}
\def\bcalL{\bar{\mathcal{L}}}
\def\bcalW{\bar{\mathcal{W}}}
\def \eqn#1#2{\begin{equation}#2\label{#1}\end{equation}}

\newcommand{\M}{\ensuremath{\mathcal{M}}
                    }
\newcommand{\oc}{\ensuremath{\overline{c}}}
\begin{titlepage}
\begin{flushright}
CHEP XXXXX
\end{flushright}
\bigskip
\def\thefootnote{\fnsymbol{footnote}}

\begin{center}
{\Large
{\bf Chiral Higher Spin Gravity \\ 
\vspace{0.1in} 
}
}
\end{center}

\bigskip
\begin{center}
{
Chethan KRISHNAN$^a$\footnote{\texttt{chethan.krishnan@gmail.com}}, Avinash RAJU$^a$\footnote{\texttt{avinashraju777@gmail.com}}}
\vspace{0.1in}

\end{center}

\renewcommand{\thefootnote}{\arabic{footnote}}

\begin{center}

$^a$ {Center for High Energy Physics,\\
Indian Institute of Science, Bangalore 560012, India}\\

\end{center}

\noindent
\begin{center} {\bf Abstract} \end{center}
We construct a candidate for the most general chiral higher spin theory with AdS$_3$ boundary conditions. In the Chern-Simons language, on the left it has the Drinfeld-Sokolov reduced form, but on the right all charges and chemical potentials are turned on. Altogether (for the spin-3 case) these are $19$ functions. Despite this, we show that the resulting metric has the form of the ``most general'' AdS$_3$ boundary conditions discussed by Grumiller and Riegler.  The asymptotic symmetry algebra is a product of a $\mathcal{W}_3$ algebra on the left and an affine $sl(3)_k$ current algebra on the right, as desired. The metric and higher spin fields depend on all the $19$ functions. We compare our work with previous results in the literature.

\vspace{1.6 cm}
\vfill

\end{titlepage}

\setcounter{footnote}{0}


\section{Introduction}

(Higher spin) gravity in 2+1 dimensions has no local propagating degrees of freedom, so the dynamics is controlled entirely  by boundary conditions. See \cite{higherspins} for an incomplete list of references on 2+1 dimensional higher spin theories in various contexts and \cite{hs-app} for their applications.

Related to this lack of local degrees of freedom is also the fact that the theory allows a Chern-Simons formulation. When there is a negative cosmological constant, it turns out that this Chern-Simons formulation contains two gauge fields, conventionally one calls these the left and right gauge fields.

In gravity theories with a negative cosmological constant, it is natural to consider Anti-deSitter space as the vacuum, and to think of solutions with asymptotically AdS$_3$ boundary conditions as states built on that vacuum. However, the precise choice of the fall-offs one allows in one's definition of ``asymptotically AdS$_3$'' has turned out to be somewhat arbitrary and many consistent choices are known in the literature \cite{refdump}. For example, the most famous of these is the Brown-Henneaux \cite{Marc} boundary conditions which corresponds in a certain radial gauge to the choice of left (right) Chern-Simons gauge field with (anti-)holomorphic dependence on the boundary coordinates, along with some restrictions on the charges and chemical potentials that show up in these gauge fields.
  
Recently, Grumiller and Riegler \cite{G-R} have written down what is arguably the ``most general'' such AdS$_3$ boundary condition for gravity. By this they mean that {\em all} the charges and chemical potentials that are visible in the Chern-Simons formulation are also visible in the metric formulation. Yet in the asymptotic limit the metric has an AdS$_3$ form, albeit with fall-offs that are more general than the ones found in Fefferman-Graham. They showed that the asymptotic symmetry algebra in this case is two copies of the $sl(2)_k$ current algebra. They accomplished this by working with a choice of radial gauge that was different from the standard radial ``Banados'' gauge. We will call this the Grumiller-Riegler radial gauge.

A very natural question to ask in this context is to see whether this can be generalized to higher spins. Can one work with a higher spin theory, turn on all the charges and chemical potentials (including higher spin ones) in the Chern-Simons language and be lead to an asymptotically AdS$_3$ metric, perhaps in the generalized Fefferman-Graham gauge of \cite{G-R}? We will see that this is (unsurprisingly) not possible. If we keep all the charges, the metric blows up at the boundary. This means that we have to remove some of the charges/chemical potentials in a consistent way\footnote{The jargon for this activity is ``reduction''.} and see whether one can get a consistent asymptotic symmetry algebra with acceptable AdS$_3$ fall-offs. In the Banados gauge, one such choice is well known: this is the so-called Drinfeld-Sokolov reduction \cite{Coussaert-Marc,Marc-Rey,Campoleoni, Campoleoni2} which when done on both sides truncates left and right sides equally and leads to an asymptotically AdS$_3$ metric in the Fefferman-Graham gauge and a $\cW_3 \times \cW_3$ asymptotic symmetry algebra. Can one allow more general charges and chemical potentials by working with the Grumiller-Riegler radial gauge and allowing the generalized Fefferman-Graham metric?

Indeed, this paper exists because we will see that if one allows this, one can do {\em much} better. In fact, one can make a Drinfeld-Sokolov reduction on only one side and (self-consistently) set a certain higher spin chemical potential to zero, while letting the other side be fully general: this still leads to the generalized Fefferman-Graham metric. This is a total count of 19 unknown functions between the charges and chemical potentials. One can also check that all of these functions show up in the metric/higher spin fields side of the story as well. Furthermore, we can calculate the asymptotic symmetry algebra and we find that the result is a copy of the $\cW_3$ algebra on the left and an affine $sl(3)_k$ algebra on the right. We suspect that this is the most general chiral higher spin gravity that satisfies these requirements, even though we do not prove it\footnote{One way to disprove this claim is to have an algebra that is ``bigger'' than $\cW_3$ but ``smaller'' $sl(3)_k$ on the left, and show that there exists a radial gauge where this, together with the right side, leads to an asymptotically AdS$_3$ fall-off. We have a suspicion that an $sl(2)_k$ on the left might also be allowed while having $sl(3)_k$ on the right, but this is a different class of chiral higher spin theory than the one we are looking at here: it does not have higher spin excitations at all on the left. Note that the right side is already as general as can be.}. 

In the final section, we will offer some comparison with the closely related results of \cite{nemani}.

\section{Chern-Simons Formulation} 

The Chern-Simons formulation of ($2+1$)-D gravity is a rewriting of the Einstein-Hilbert action with or without a cosmological constant into a Chern-Simons gauge theory with appropriate gauge group (see eg. \cite{GrassmannPath}). When the cosmological constant is negative

\begin{equation}
S_{\mathrm{Grav.}} = I_{CS}[A] - I_{CS}[\bA]
\end{equation}
where 

\begin{equation}
I_{CS}[A] = \frac{k}{4\pi}\int_{\calM}\bra A \wedge \mathrm{d}A + \frac{2}{3}A\wedge A\wedge A \ket.
\end{equation}
Here $k=\frac{l}{4G_N}$ is the Chern-Simons level and $l$ is the AdS radius and $G_N$ is the Newton's constant. The appropriate gauge group is decided by the sign of the cosmological constant and for a negative cosmological constant $A$ and $\bA$ takes values in $sl(2,R)$ Lie algebra. The connections $A$ and $\bA$ are related to the the veilbein $e_{\mu}^a$ and spin connection $\omega^{a}_{\mu} = \e^{abc}\omega_{bc \mu}$ as follows

\begin{equation}\label{connections}
A = \left(\omega^{a}_{\mu}L_a + \frac{e^{a}_{\mu}}{l}L_a \right)dx^{\mu}, \quad \bA = \left(\omega^{a}_{\mu}L_a - \frac{e^{a}_{\mu}}{l}L_a \right)dx^{\mu}.
\end{equation}
The equations of motion gives the flatness condition $F = \bar{F} =0$ where

\begin{equation}
F= \mathrm{d}A + A\wedge A =0
\end{equation}
and is equivalent to Einstein's equation. $L_a$ is the generator of $sl(2,R)$ on which is defined the invariant bilinear form

\begin{equation}
\bra L_a L_b \ket = \frac{1}{2}\eta_{ab}.
\end{equation}
Using eq. \eqref{connections}, $A$ and $\bA$ is related to the metric through the veilbein $e = \frac{1}{2}(A- \bA)$ and 

\begin{equation}\label{metric}
g_{\mu \nu} = \frac{1}{2}\bra e_{(\mu} e_{\nu)} \ket
\end{equation}
In $2+1$ dimensions, the Chern-Simons formulation lends itself to generalization to higher spin theories coupled to gravity by lifting the gauge group to $sl(n,R)$. An $sl(n,R)$ Chern-Simons theory describes gravity coupled to a tower of spins $3,4,\cdots,n$. In metric-like formulation, a spin-$j$, ($j \leq n$) field is given by 

\begin{equation}
\varphi_{\mu_1 \cdots \mu_j} = \frac{1}{j!}\bra e_{(\mu_1}\cdots e_{\mu_j)} \ket
\end{equation}

\section{Chiral Spin-3 Gravity}

Taking a cue from the empty AdS solution, we can choose a radial gauge of the form

\begin{equation}\label{radial_bc}
A = b^{-1}\mathrm{d}b + b^{-1}a(t,\phi)b, \quad \bA = b\mathrm{d}b^{-1} + b\ba(t,\phi)b^{-1}
\end{equation}
where all the $\rho$ dependence now comes from the group element $b(\rho)$. For empty AdS there are many choices of radial gauge, the usual one being $b = \exp(\rho L_0)$, which we will call the Banados gauge. In \cite{G-R}, the group element $b$ was instead chosen to be

\begin{equation}\label{radial_gauge}
b= \exp(L_{-1})\exp(\rho L_0)
\end{equation}
This radial gauge (which we will call the Grumiller-Riegler gauge) manifests all the $sl(2)$ charges and potentials present in the gauge field language, in the metric language as well. For our case, $a$ and $\ba$ are the $sl(3,R)$ Lie algebra valued fields which takes the general form

\begin{eqnarray}\label{gen_connection}
a_t(t,\phi) &=& \mui{-1}L_{-1} + \mui{0}L_0 + \mui{1}L_{1} + \nui{-2}W_{-2} + \nui{-1}W_{-1} \br &\quad \quad &+ \nui{0}W_0 + \nui{1}W_1 + \nui{2}W_2 \br 
a_{\phi}(t,\phi) &=&  \calL{-1}L_{-1} + \calL{0}L_0 + \calL{1}L_1 + \calW{-2}W_{-2} + \calW{-1}W_{-1} \br &\quad \quad &  +\; \calW{0}W_0 + \calW{1}W_1 + \calW{2}W_2 \br
\ba_t(t,\phi) &=& \bmu^{(-1)}L_{-1} + \bmu^{(0)}L_{0} + \bmu^{(1)}L_{1} + \bnu^{(-2)}W_{-2} + \bnu^{(-1)}W_{-1} \br &+& \bnu^{(0)}W_{0} + \bnu^{(1)}W_{1} + \bnu^{(2)}W_{2}\\ 
\ba_{\phi}(t,\phi) &=&  \frac{\pi}{8k}\left( \bcalL^{(-1)} L_{-1} - 2\bcalL^{(0)} L_{0} +  \bcalL^{(1)} L_{1} - \frac{1}{4} \bcalW^{(-2)}W_{-2} + \bcalW^{(-1)}W_{-1} \right. \br &\quad \quad & - \; \left. \frac{3}{2} \bcalW^{(0)}W_{0} + \bcalW^{(1)}W_{1} - \frac{1}{4} \bcalW^{(-2)}W_{-2}  \right) \nonumber
\end{eqnarray}
where $\{L_i,W_j \}$ are the $sl(3,R)$ generators whose algebra is given by \eqref{generators}. The asymmetry between the left and right normalizations arises because on the left we are eventually going to work with the Drinfeld-Sokolov reduced form. The normalization of coefficients of $\ba_{\phi}$ have been chosen so that our expression for the global charge, \eqref{sl3_charge}, comes out in the canonical form. The equations of motion impose relationship between the set of potentials $\{\mui{i}, \nui{j} \}$ and the charges $\{\calL{i}, \calW{j} \}$ and is given in the appendix for the unbarred sector while it is identical for the barred sector also. Following \cite{G-R}, we hold chemical potentials as fixed functions $\delta a_t = \delta \ba_t =0$ while

\begin{eqnarray}
\delta a_{\phi}(t,\phi) &=& \sum_{i=-1}^{1} \delta \calL{i}(t,\phi)L_i + \sum_{j=-2}^{2} \delta \calW{j}(t,\phi)W_j \br
\delta \ba_{\phi}(t,\phi) &=& \sum_{i=-1}^{1} \delta \bcalL{i}(t,\phi)L_i + \sum_{j=-2}^{2}\delta \bcalW{j}(t,\phi)W_j \br
\end{eqnarray}
The time-like boundary of AdS gives rise to an infinite dimensional phase space with infinitely many global charges. The algebra of charges can be then determined by the Regge-Teitelboim approach \cite{Regge,Banados}.

In \cite{G-R}, a generalization of Fefferman-Graham gauge for asymptotically AdS metrics was introduced. The motivation of this gauge was to capture all independent $sl(2)$ charges in the metric formulation as well. This metric takes the general form

\begin{eqnarray}\label{ads_metric}
ds^2 &=& d\rho^2 + 2\left(e^{\rho}N^{(0)}_i + N^{(1)}_i + e^{-\rho}N^{(2)}_i + O(e^{-2\rho}) \right)d\rho dx^i \br &+& \left( e^{2\rho}g^{(0)}_{ij} + e^{\rho }g^{(1)}_{ij} + g^{(2)}_{ij} + O(e^{-\rho}) \right)dx^i dx^j
\end{eqnarray}
We call this the \textit{generalized Fefferman-Graham} gauge and various expansion coefficients capture all independent combinations of the chemical potentials and charges in the $sl(2)$ Chern-Simons language. This was the motivation for the choice of radial gauge \eqref{radial_gauge}. 

We take \eqref{ads_metric} as our definition of \textit{asymptotically AdS} metric and construct a higher spin theory which preserves this form of metric. Using \eqref{radial_gauge}, \eqref{radial_bc} and \eqref{metric}, we find that the most general gauge connection \eqref{gen_connection} would violate the metric form \eqref{ads_metric} and therefore the coefficients of the gauge connection have to be restricted. We begin by imposing the Drinfeld-Sokolov (DS) condition on the left gauge connection $a_{\phi}$ in order to further restrict the coefficients. DS reduction in the highest weight gauge amounts to fixing the coefficients of $a_{\phi}$ such that 

\begin{equation}
a_{\phi}(\phi) = L_1 + a^{(-)}_{\phi}(\phi)
\end{equation}
where 

\begin{equation}
[L_{-1}, a^{(-)}_{\phi}(\phi)] = 0
\end{equation}
The above equation fixes the form of $a^{(-)}$ and the reduced gauge field can be now written as

\begin{equation}
a_{\phi}(\phi) = L_1 + \frac{2\pi}{k}{\cal L}L_{-1} -\frac{\pi}{2k} {\cal W}W_{-2}
\end{equation}
As shown by \cite{Campoleoni, Campoleoni2}, this restriction helps getting a $\cW_3$ charge algebra. With this restriction on $a_{\phi}$, the equations of motion takes the form

\begin{eqnarray}
\mui{0} + \pp \mui{1} &=& 0 \br 
\cL \mui{1} + 2\cW \nui{2} - \frac{k}{2\pi} \mui{-1} - \frac{k}{4\pi}\pp \mui{0} &=& 0 \br 
\cL \mui{0} + \cW \nui{1} - \frac{k}{2\pi} \pp \mui{-1} + \pt \cL &=& 0 \br 
2\cL \nui{-1} - \cW \mui{0} - \frac{k}{\pi}\pp \nui{-2} - \frac{1}{2}\pt \cW &=& 0 \br 
\cL \nui{0} - \frac{1}{2}\cW \mui{1} - \frac{k}{\pi}\nui{-2} - \frac{k}{4\pi}\pp \nui{-1} &=& 0 \\ 
\cL \nui{1} - \frac{k}{2\pi}\nui{-1} - \frac{k}{6\pi}\pp \nui{0} &=& 0 \br 
\cL \nui{2} - \frac{k}{4\pi}\nui{0} - \frac{k}{8\pi}\pp \nui{1} &=& 0 \br 
\nui{1} + \pp \nui{2} &=& 0 \nonumber
\end{eqnarray}
However, to get the right fall-offs for the metric it turns out that the DS reduction is not enough. But it can be accomplished by specifying the chemical potentials. This is consistent, because chemical potentials are fixed functions that we are allowed to specify. For the metric to be have correct fall-off we need $\nui{2} =0$. This fact combined with the equations of motion (partially) fix other chemical potentials. It turns out that the chemical potentials are now specified by a single function $\mu$. The final left gauge connection can be written as

\begin{eqnarray}\label{gauge_sol_unbarred}
a &=& \left(\mu L_1 - \mu' L_0 + \left( \frac{2\pi}{k}{\cal L}\mu +\frac{1}{2}\mu'' \right)L_{-1} -\frac{\pi}{2k} {\cal W}\mu W_{-2} \right)dt \br &+& \left( L_1 + \frac{2\pi}{k}{\cal L}L_{-1} - \frac{\pi}{2k}{\cal W} W_{-2} \right)d\phi \\
\end{eqnarray}
and the left over equations of motion for $a(t,\phi)$ now gives

\begin{eqnarray}
\dot{\cL} - \mu \cL' - 2\cL \mu' - \frac{k}{4\pi}\mu''' &=& 0 \\
-\dot{\cW} + \mu \cW' + 3\cW \mu' &=& 0
\end{eqnarray}
where dot and prime refers to derivatives with respect to time and $\phi$ respectively. Note that setting $\mu=1$ results in holomorphic dependence of ${\cal L}$ and ${\cal W}$ on the boundary coordinates \cite{Campoleoni}.

For the barred sector, no further restrictions are needed to stay in the generalized Fefferman-Graham form of the metric, and therefore $\ba(t,\phi)$ stays in the most general form:

\begin{eqnarray}\label{gauge_sol_barred}
\ba(t,\phi) &=& \left( \bmu^{(-1)}L_{-1} + \bmu^{(0)}L_{0} + \bmu^{(1)}L_{1} + \bnu^{(-2)}W_{-2} + \bnu^{(-1)}W_{-1} \right. \br &+& \left. \bnu^{(0)}W_{0} + \bnu^{(1)}W_{1} + \bnu^{(2)}W_{2}\right) dt + \frac{\pi}{8k}\Bigg( \bcalL^{(-1)} L_{-1} - 2\bcalL^{(0)} L_{0} +  \bcalL^{(1)} L_{1} \br  &-& \frac{1}{4} \bcalW^{(-2)}W_{-2} + \bcalW^{(-1)}W_{-1} - \frac{3}{2} \bcalW^{(0)}W_{0} + \bcalW^{(1)}W_{1} - \frac{1}{4} \bcalW^{(-2)}W_{-2}  \Bigg) \nonumber
\end{eqnarray}
The equations of motion for $\ba(t,\phi)$ is substantially more complicated because of more number of terms and is listed in the appendix (where we suppress the bar's however). The metric is explicitly given by: 

\begin{eqnarray} \label{full_metric}
g_{\rho \rho} &=&  1, \quad g_{\rho t} = \frac{\rme^{\rho}}{2}\mu + \left( \mu - \frac{1}{2}\bmu^{(0)} - \frac{1}{2}\mu' \right) + \frac{\rme^{-\rho}}{2}\bmu^{(1)} \br 
g_{\rho \phi} &=& \frac{\rme^{\rho}}{2} + \left( 1- \frac{\pi}{8k}\bcalL^{(0)} \right) +  \frac{\rme^{-\rho}\pi}{16k}\bcalL^{(1)} \br 
g_{tt} &=& \rme^{2\rho} \mu \bmu^{(-1)} - \rme^{-\rho} \mu \bmu^{(0)} + \left[ -\frac{2\pi}{k}\cL \mu^2 - \mu\bmu^{(0)} + \frac{1}{4}\bmu^{(0)2} + \mu \bmu^{(1)}  - \bmu^{(1)}\bmu^{(1)} + \frac{1}{3}\bnu^{(0)2} \right. \br &-& \left. \bnu^{(1)}\bnu^{(-1)} + 4\bnu^{(2)}\bnu^{(-2)} + \frac{1}{2}\bmu^{(0)}\mu' + \frac{1}{4}\mu'^{2} - \frac{1}{2}\bmu^{(1)}\mu''  \right] + \rme^{-\rho} \bmu^{(1)}(2\mu - \mu') \br &+& \rme^{-2\rho}\left( \bmu^{(1)}\mu + \frac{2\pi}{k}\cL \mu \bmu^{(1)} - \bmu^{(1)}\mu' + \frac{1}{2}\bmu^{(1)}\mu''   \right) + \rme^{-4\rho}\frac{2\pi }{k}\cW \mu \bnu^{(2)} \br
g_{t\phi} &=& \frac{\rme^{2\rho}}{2}\left( \frac{1}{4}\bcalL^{(-1)}\mu + \bmu^{(-1)} \right) - \frac{\rme^{\rho}}{2}\left( \frac{1}{4}\bcalL^{(0)}\mu - \bmu^{(0)}   \right) + \frac{\pi}{16k}\left[ - 32\cL \mu + 2\bcalL^{(0)}\mu \right. \br &+& \left. \bcalL^{(1)}\mu - \frac{8k}{\pi}\bmu^{(0)} - \bcalL^{(0)}\bmu^{(0)} + \frac{8k}{\pi}\bmu^{(1)} - \bcalL^{(-1)}\bmu^{(1)} -\bcalL^{(1)}\bmu^{(-1)}- \bcalW^{(0)}\bnu^{(0)} \right. \br &-& \left. \bcalW^{(-1)}\bnu^{(1)} - \bcalW^{(-2)}\bnu^{(2)} - \bcalW^{(1)}\bnu^{(-1)} - \bcalW^{(2)}\bnu^{(-2)} - \bcalL^{(0)}\mu' - \frac{4k}{\pi}\mu'' \right] \br &+& \frac{\rme^{-\rho}\pi}{2k} \left[\frac{2k}{\pi}\bmu^{(1)} + \frac{\bcalL^{(1)}}{4}\left(\mu - \frac{1}{2}\mu' \right)   \right] + \frac{\rme^{-2\rho}\pi}{4k}\left[ \frac{1}{4}\bcalL^{(1)}\mu + \frac{\pi}{2k}\bcalL^{(1)}\cL \mu + \frac{2k}{\pi}\bmu^{(1)} \right. \br &+& \left. 4\cL\bmu^{(1)} - \frac{1}{4}\bcalL^{(1)}\mu' + \frac{1}{8}\bcalL^{(1)}\mu''  \right] + \rme^{-4\rho}\frac{\pi}{k}\cW \left(  \bnu^{(2)} - \frac{\pi}{32k}\bcalW^{(2)}\mu  \right) \\ 
g_{\phi \phi} &=& \rme^{2\rho}\frac{\pi}{8k}\bcalL^{(-1)} + \rme^{\rho}\frac{\pi}{4k}\bcalL^{(0)} + \frac{\pi}{4k}\left[ \bcalL^{(0)}+ \frac{\pi}{16k}\bcalL^{(0)2} + \frac{1}{2}\bcalL^{(1)} - \frac{\pi}{16k}\bcalL^{(1)}\bcalL^{(-1)} \right. \br &-& \left.  8\cL + \frac{3\pi}{64k}\bcalW^{(0)2} -\frac{\pi}{16k}\bcalW^{(1)}\bcalW^{(-1)} + \frac{\pi}{16k}\bcalW^{(2)}\bcalW^{(-2)} \right] + \rme^{-\rho}\frac{\pi}{4k}\bcalL^{(1)} \br &+& \rme^{-2\rho}\frac{\pi}{8k^2}\bcalL^{(1)}\left( k + 2\pi \cL \right) - \rme^{-4\rho}\frac{\pi^2}{16k^2}\bcalW^{(2)}\cW \nonumber
\end{eqnarray}
We omit writing the spin-3 field here to avoid clutter. However we like to remark that the metric and the spin-3 field combined has all the 19 independent functions that appeared in the gauge field. In the next section, we present the asymptotic symmetry algebra for our solution.

\section{Global Charges and their Algebra}
In this section we elucidate the approach of Regge-Teitelboim \cite{Regge}, applied to Chern-Simons theories \cite{Banados} to compute the algebra of global charges. Asymptotically AdS spaces have a time-like boundary and the gauge transformations that act non-trivially on the boundary (\textit{i.e.} gauge transformations that do not become identity as $\rho \rightarrow \infty$) are not true gauge transformations but are rather genuine symmetry transformations. On the time-like boundary, these transformations map one solution to another giving rise to a non-trivial boundary phase space.

We begin with a space + time decomposition of the CS gauge field\footnote{Our discussion in this section is for the pure Chern-Simons theory with no extra boundary terms. This corresponds to a Neumann boundary condition in the metric language, see \cite{Neumann}. The variational principle with $a_t$ held fixed that we have used in the previous sections requires the addition of an additional boundary term to the CS action, to be well-defined. But this point will not affect our discussion in this section because it relies only on the Lagrange multiplier term that generates the gauge algebra, which arises from the bulk piece in \eqref{lagrange}.}. We begin by assuming that the three manifold $\calM$ is topologically a solid cylinder with a time-like boundary which is topologically $\partial \calM \simeq \mathbb{R}\times S^1$. The gauge connection can be expressed as 

\begin{equation}
A_{\mu}dx^{\mu} = A_t dt + A_{\rho} d\rho + A_{\phi} d\phi
\end{equation}
The action now takes the form

\begin{equation}
I_{CS} = \frac{k}{4\pi}\int_{\calM}dt d\rho d\phi \bra A_{\phi}\dot{A}_{\rho} - A_{\rho}\dot{A}_{\phi} + 2A_t F_{\rho \phi} \ket + \frac{k}{4\pi}\int_{\partial \calM}dt d\phi \bra A_t A_{\phi} \ket \label{lagrange}
\end{equation}
In the above action we see that $A_t$ is a Lagrange multiplier enforcing the constraint $F_{\rho \phi}=0$ while $A_{\rho}$ and $A_{\phi}$ are the dynamical variables (with the caveat that this theory is topological). Similar story will also hold for the barred sector $\bA_{\mu}$. We first define the smeared generators of gauge transformations

\begin{equation}
G[\Lambda] = \frac{k}{4\pi}\int_{\Sigma}dtd\phi \bra \Lambda F_{t\phi} \ket + Q[\Lambda]
\end{equation} 
where $\Sigma$ is the spatial hypersurface and $Q[\Lambda]$ is a boundary term that is added to make $G[\Lambda]$ a differentiable functional of $A_i, i=\rho, \phi$. For a state independent gauge parameter $\Lambda$, $Q[\Lambda]$ takes the form

\begin{equation}
Q[\Lambda] = -\frac{k}{2\pi} \int_{\partial \Sigma}d\phi \bra \Lambda A_{\phi} \ket
\end{equation}
The generators $G(\Lambda)$ satisfies the Poisson bracket relation

\begin{equation}\label{G-algebra}
\{ G[\Lambda], G[\Gamma] \} = G[[\Lambda, \Gamma]] + \frac{k}{2\pi}\int_{\partial \Sigma}d\phi \bra \Lambda \partial_{\phi}\Gamma \ket
\end{equation}
The second term is a central extension and comes out as a consequence of the surface term $Q[\Lambda]$. $Q[\Lambda]$ does not vanish on-shell, \textit{i.e.} when $F_{ij}=0$ and any transformation for which $Q[\Lambda]\neq 0$ generate global symmetries through 

\begin{equation}\label{gauge_var}
\delta_{\Lambda}F = \{Q[\Lambda],F[A_i] \}
\end{equation}
for any phase space functional $F[A_i]$. The gauge transformations $\Lambda$ that preserve the boundary condition \ref{radial_bc} and \ref{gen_connection} can be written as

\begin{equation}\label{gauge_param_1}
\Lambda = b^{-1}\lambda(t,\phi) b, \quad \lambda = \ei{i}(t,\phi)L_i + \si{j}(t,\phi)W_j
\end{equation}
where the summation is implied over indices $i$ and $j$. This gives a non-vanishing charge $Q[\Lambda]$

\begin{equation}\label{surface_charge}
Q[\Lambda] = -\frac{k}{2\pi}\int d\phi \bra \lambda a_{\phi} \ket
\end{equation}

Now we are in a position to compute the Poisson algebra of the charges. We begin with the unbarred sector whose coefficients are restricted by the DS highest weight gauge condition. Under the gauge variation \ref{gauge_param_1}, the connection $a$ transforms by

\begin{equation}
\delta_{\lambda}a = {\rm d}\lambda + [a,\lambda]
\end{equation}

Since the form of the gauge field is fixed, the boundary condition preserving transformations $\Lambda$, given by \eqref{gauge_param_1},  are now characterized by two functions $\e$ and $\sigma$. The exact form of the gauge transformation parameter and variation of charges can be found in \cite{Campoleoni}. The expression for the global charge associated with the transformation is given by

\begin{equation}
Q[\Lambda] = \oint d\phi \left( \e(\phi){\cal L}(\phi) + \sigma(\phi){\cal W}(\phi) \right)
\end{equation}
The charges generate an algebra through the Poisson brackets which is nothing but the classical ${\cal W}_3$ algebra 

\begin{eqnarray}
\{{\cal L}(\phi), {\cal L}(\phi') \} &=& -\left( \delta(\phi-\phi'){\cal L}'(\phi) + 2\delta'(\phi-\phi'){\cal L}(\phi) \right) - \frac{k}{4\pi}\delta'''(\phi-\phi') \\
\{{\cal L}(\phi), {\cal W}(\phi') \} &=& -\left( 2\delta(\phi-\phi'){\cal W}'(\phi) + 3\delta'(\phi-\phi'){\cal W}(\phi) \right) \\
\{{\cal W}(\phi), {\cal W}(\phi') \} &=& -\frac{1}{3} \left( 2\delta(\phi-\phi'){\cal L}'''(\phi) + 9\delta'(\phi-\phi'){\cal L}''(\phi) + 15\delta''(\phi-\phi'){\cal L}'(\phi)\right. \br &+& \left. 10\delta'''(\phi-\phi'){\cal L}(\phi) + \frac{k}{4\pi}\delta^{(5)}(\phi-\phi') \right. \br &+& \left. \frac{64\pi}{k}\left( \delta(\phi-\phi'){\cal L}{\cal L}'(\phi) + \delta'(\phi-\phi'){\cal L}^2(\phi)  \right)   \right)
\end{eqnarray}
with the central charge

\begin{equation}
c= 6k = \frac{3l}{2G_N}
\end{equation}
In terms of the Fourier modes

\begin{equation}
{\cal L}(\phi) = - \frac{1}{2\pi}\sum L_n {\rm e}^{-in\phi}, \quad {\cal W}(\phi) =  \frac{1}{2\pi}\sum W_n {\rm e}^{-in\phi}
\end{equation}
and shifting the zero mode of ${\cal L}$ by

\begin{equation}
L_0 \rightarrow L_0 - \frac{k}{4}.
\end{equation}
The algebra now takes the familiar form

\begin{eqnarray}
i\{ L_m, L_n \} &=& (m-n)L_{m+n} + \frac{c}{12}\delta_{m+n,0} \\
i\{ L_m, W_n \} &=& (2m-n)W_{m+n} \\
i\{ W_m, W_n \} &=& -\frac{1}{3} \left( (m-n)(2m^2 + 2n^2 -mn -8)L_{m+n} + \frac{96}{c}(m-n)\Lambda_{m+n} \right. \br &+& \left. \frac{c}{12}m(m^2 -1)(m^2 -4)\delta_{m+n,0} \right)
\end{eqnarray}
where $\Lambda_n$ is defined as

\begin{equation}
\Lambda_n = \sum_{n \in \mathbb{Z}}L_{m+n}L_{-m}
\end{equation}

Now we return to the barred sector and compute the symmetry algebra. Since there are no restrictions on the coefficients of $\ba_{\phi}$ all the charges transform under the gauge transformation. The boundary preserving gauge transformation is given by
 \begin{equation}\label{gauge_param}
\overline{\Lambda} = b \overline{\lambda}(t,\phi) b^{-1}, \quad \overline{\lambda} = \bei{i}(t,\phi)L_i + \bsi{j}(t,\phi)W_j
\end{equation}
The transformation of charges under the above gauge transformation is then given by

\begin{eqnarray}\label{charge_var}
\delta \bcalL^{(0)} &=& \frac{1}{4}\Bigg(\bcalL^{(-1)}\bei{1} -\bcalL^{(1)}\bei{-1}+ \bcalW^{(-1)}\bsi{1} +2\bcalW^{(-2)}\bsi{2} - \bcalW^{(1)}\bsi{-1} \br && \hspace*{5cm} -2\bcalW^{(2)}\bsi{-2}  -\frac{k}{\pi}\partial_{\phi}\bei{0} \Bigg) \br
\delta \bcalL^{(-1)} &=& \frac{1}{4} \Bigg(-\bcalL^{(-1)}\bei{0} - 2\bcalL^{(0)}\bei{-1} - 2\bcalW^{(-1)}\bsi{0} - \bcalW^{(-2)}\bsi{1} - 3\bcalW^{(0)}\bsi{-1} \br && \hspace*{5cm} - 4\bcalW^{(1)}\bsi{-2} + \frac{2k}{\pi}\partial_{\phi}\bei{-1} \Bigg) \br
\delta \bcalL^{(1)} &=& \frac{1}{4} \Bigg( \bcalL^{(1)}\bei{0} + 2\bcalL^{(0)}\bei{1} + 2\bcalW^{(1)}\bsi{0} + 3\bcalW^{(0)}\bsi{1} + 4\bcalW^{(-1)}\bsi{2} \br && \hspace*{5cm} + \bcalW^{(2)}\bsi{-1} + \frac{2k}{\pi}\partial_{\phi}\bei{1} \Bigg) \br
\delta \bcalW^{(2)} &=& \frac{1}{4}\Bigg( 2\bcalW^{(2)}\bei{0} + 4\bcalW^{(1)}\bei{1} - 4\bcalL^{(1)}\bsi{1} - 16\bcalL^{(0)}\bsi{2} - \frac{8k}{\pi}\partial_{\phi}\bsi{2} \Bigg) \\
\delta \bcalW^{(1)} &=& \frac{1}{4} \Bigg( \bcalW^{(1)}\bei{0} + 3\bcalW^{(0)}\bei{1} - \bcalW^{(2)}\bei{-1}  + 2\bcalL^{(1)}\bsi{0} + 2\bcalL^{(0)}\bsi{1} \br && \hspace*{5cm} - 4\bcalL^{(-1)}\bsi{2} + \frac{2k}{\pi}\partial_{\phi}\bsi{1} \Bigg) \br
\delta \bcalW^{(0)} &=& \frac{1}{4}\Bigg( 2\bcalW^{(-1)}\bei{1} - 2\bcalW^{(1)}\bei{-1} + 2\bcalL^{(-1)}\bsi{1} - 2\bcalL^{(1)}\bsi{-1} - \frac{4k}{3\pi}\partial_{\phi}\bsi{0} \Bigg)\br
\delta \bcalW^{(1)} &=& \frac{1}{4}\Bigg(-\bcalW^{(-1)}\bei{0} + \bcalW^{(-2)}\bei{1} - 3\bcalW^{(0)}\bei{-1} - 2\bcalL^{(-1)}\bsi{0} - 2\bcalL^{(0)}\bsi{-1} \br && \hspace*{5cm} + 4\bcalL^{(1)}\bsi{-2} + \frac{2k}{\pi}\partial_{\phi}\bsi{-1} \Bigg) \br
\delta \bcalW^{(2)} &=& \frac{1}{4}\Bigg(-2\bcalW^{(-2)}\bei{0} - 4\bcalW{(-1)}\bei{-1} + 4\bcalL^{(-1)}\bsi{-1} + 16\bcalL^{(0)}\bsi{-2} - \frac{8k}{\pi}\partial_{\phi}\bsi{-2} \Bigg) \nonumber
\end{eqnarray}
Similarly, the $\ba_t$ is fixed, $\delta_{\lambda}\ba_t = 0$ gives the time-evolution of the gauge parameters $\{\bei{i},\bsi{j} \}$. From \eqref{surface_charge}, the charge associated with the above gauge transformation can be computed to be

\begin{eqnarray}\label{sl3_charge}
\bar{Q}[\overline{\lambda}] &=& \oint d\phi \left(\bcalL^{(0)}\bei{0} + \bcalL^{(1)}\bei{-1} + \bcalL^{(-1)}\bei{1} + \bcalW^{(2)}\bsi{-2} + \bcalW^{(1)}\bsi{-1}  + \bcalW^{(0)}\bsi{0} \right. \br  &+& \left. \bcalW^{(-1)}\bsi{1} + \bcalW^{(-2)}\bsi{2}  \right)
\end{eqnarray}
Using eq. \eqref{gauge_var} and eq. \eqref{charge_var}, the Poisson algebra of the connections can be determined to be an $sl(3)_{k}$ Kac-Moody algebra \cite{Campoleoni}.

\begin{equation}
\{ \ba^{A}_{\phi}(\phi),\ba^{B}_{\phi}(\phi') \} = -\frac{2\pi}{k}\left(\delta(\phi - \phi')f^{AB}_{\;\;\;\;\; C}\ba^{C}_{\phi}(\phi) - \delta'(\phi-\phi')\gamma^{AB}  \right) .
\end{equation}
$f^{AB}_{\;\;\;\;\; C}$ are the structure constants of $sl(3)$ and $\gamma^{AB}$ is the inverse of the $sl(3)$ Killing metric. The $sl(3)_k$ algebra can be written in a more familiar form by Fourier decomposing the gauge connection 

\begin{eqnarray}
\ba^{A}(\phi) = \frac{1}{k}\sum_{n \in \mathbb{Z}}\ba^{A}_n \mathrm{e}^{-i n\phi}
\end{eqnarray}
which gives

\begin{equation}
\{ \ba^{A}_n, \ba^{B}_m \} = -f^{AB}_{\;\;\;\;\; C} \ba^{C}_{n+m} + in\gamma^{AB}\delta_{n+m,0}
\end{equation}
Thus our solution presents a $\cW_3 \times sl(3)_k$ as its asymptotic symmetry algebra.

\section{Comments}

In this concluding section, we briefly contrast our work with previous results. In an interesting paper \cite{nemani} Poojary and Suryanarayana made the following choice of the bare gauge field:
\begin{eqnarray}\label{nemani_gauge1}
a &=& \left( L_1 - \kappa L_{-1} - \omega W_{-2} \right)dt + \left( L_1 - \kappa L_{-1} - \omega W_{-2} \right)d\phi \\
\ba &=& \left(  -L_{-1} + \tilde{\kappa} L_{1} + \tilde{\omega} W_{2} + \sum^{1}_{a=-1}f^a L_a + \sum_{b=-2}^{2} g^b W_b \right)dt \br &+& \left( L_{-1} - \tilde{\kappa} L_{1} - \tilde{\omega} W_{2} + \sum^{1}_{a=-1}f^a L_a + \sum_{b=-2}^{2} g^b W_b \right)d\phi\label{nemani_gauge2}
\end{eqnarray}
In our language, this amounts to the following restrictions on the charges and the chemical potential: 
\begin{eqnarray}
\mu &=& 1,\;\; \cL = -\frac{k}{2\pi}\kappa,\;\; \cW = \frac{2k}{\pi}\omega,\;\; \br 
\bmu^{(1)} &=& \tilde{\kappa} + f^{1},\;\; \bmu^{(-1)} = f^{-1} -1,\;\;\bmu^{(0)} = f^0 \\ 
\bnu^{(2)} &=& g^2 + \tilde{\omega},\quad \bnu^{(a)} = \bcalW^{(a)} = g^a, a\neq 2 \br 
\bcalL^{(1)} &=& \frac{8k}{\pi}(f^1 - \tilde{\kappa}),\;\; \bcalL^{(-1)} = \frac{8k}{\pi}(1+f^{-1}), \;\; , \bcalL^{(0)} = -\frac{4k}{\pi}f^0, \;\; \br \bcalW^{(2)} &=& -\frac{32k}{\pi}(g^2 - \tilde{\omega}),\;\; \bcalW^{(-2)} = -\frac{32k}{\pi}g^{-2}, \;\; \bcalW^{(0)} = -\frac{16k}{3\pi} g^0, \br \bcalW^{(\pm1)} &=& \frac{8k}{\pi}g^{\pm 1}
\end{eqnarray}
Note that there are 12 independent functions in this case as opposed to our 19. Furthermore imposing the equations of motion makes $\kappa$ and $\omega$ holomorphic in the boundary coordinates. They showed that with this choice the asymptotic charge algebra is $\cW_3 \times sl(3)_k$. 

Since the charge algebra depends only on the bare gauge field, our construction is an explicit demonstration that the restriction of \cite{nemani} is not necessary if one's goal is to reproduce the $\cW_3 \times sl(3)_k$ algebra: the most general gauge field on the right side, together with a somewhat more general gauge field on the left (see Section 3 for details), will still do the job. 


Another comment worth making is that even with the restricted form \eqref{nemani_gauge1}, \eqref{nemani_gauge2} of \cite{nemani}, one can check that the metric does {\em not} have the typical Fefferman-Graham fall-off, neither in the Banados radial gauge that \cite{nemani} are working with (we show this in an Appendix), nor in the Grumiller-Riegler radial gauge that we use (which is a corollary of our results in section 3).  To make sense as an asymptotically AdS space, one {\em must} think of the metric in the generalized Fefferman-Graham gauge\footnote{This issue is possibly moot however if one views asymptotic AdS$_3$ in higher spin theories as a not-very-meaningful idea in the metric formulation. If one adopts such a draconian point of view, there is nothing stopping one from turning on all the charges on either side and the resulting charge algebra would be two copies of $sl(3)_k$. It should be kept in mind however, that part of the motivation for the Drinfeld-Sokolov reduction choice in \cite{Campoleoni} was that the metric had the usual (asymptotic) AdS$_3$ form even with higher spins turned on. This is our motivation for taking the metric (somewhat) seriously even with higher spins turned on.}. We have checked that our general field configuration leads to this gauge in both Banados and Grumiller-Riegler radial gauges.


\section*{Acknowledgement}

We thank Andrea Campoleoni for a clarifying correspondence and Max Riegler for comments on this manuscript. CK thanks Shubho Roy for a preliminary collaboration on the possibility of a chiral higher spin gravity back in 2013-14, and Nemani Suryanarayana for discussions. 

\appendix
\section{$sl(3)$ algebra and generators}
For $sl(3)$ generators we use the principle embedding basis where the generators $\{ L_i, W_j \}$ satisfy the algebra

\begin{equation}
[L_i,L_j]  =  (i-j) L_{i+j}
\end{equation}
\begin{equation}
[L_i,W_j]  =  (2i-j) W_{i+j}
\end{equation}
\begin{equation}
[W_i, W_j] = -\frac{1}{3}(i-j)(2i^2 + 2j^2 -ij -8)L_{i+j}
\end{equation}

We work with the $3\times 3$ fundamental representation and the generators are explicitly given by

\begin{eqnarray}\label{generators}
L_{-1} &=& \left(
\begin{array}{ccc}
 0 & -2 & 0 \\
 0 & 0 & -2 \\
 0 & 0 & 0 \\
\end{array}
\right), \quad L_0 = \left(
\begin{array}{ccc}
 1 & 0 & 0 \\
 0 & 0 & 0 \\
 0 & 0 & -1 \\
\end{array}
\right), \quad L_1 = \left(
\begin{array}{ccc}
 0 & 0 & 0 \\
 1 & 0 & 0 \\
 0 & 1 & 0 \\
\end{array}
\right) \br 
W_{-2} &=& \left(
\begin{array}{ccc}
 0 & 0 & 8 \\
 0 & 0 & 0 \\
 0 & 0 & 0 \\
\end{array}
\right), \quad  W_{-1} = \left(
\begin{array}{ccc}
 0 & -2 & 0 \\
 0 & 0 & 2 \\
 0 & 0 & 0 \\
\end{array}
\right), \quad W_0 = \frac{2}{3}\left(
\begin{array}{ccc}
 1 & 0 & 0 \\
 0 & -2 & 0 \\
 0 & 0 & 1 \\
\end{array}
\right) \\
W_1 &=& \left(
\begin{array}{ccc}
 0 & 0 & 0 \\
 1 & 0 & 0 \\
 0 & -1 & 0 \\
\end{array}
\right), \quad W_2 = \left(
\begin{array}{ccc}
 0 & 0 & 0 \\
 0 & 0 & 0 \\
 2 & 0 & 0 \\
\end{array}
\right) \nonumber
\end{eqnarray} 
The Killing metric of $sl(3)$ is given by 

\begin{equation}
\gamma_{AB} = x\; \Tr[T_A,T_B], \quad x \in \mathbb{R}
\end{equation}
where $T_A \in \{L_i,W_j \},\; i=-1,\cdots,1,\;j=-2,\cdots,2$. We choose the constant $x$ such that

\begin{equation}
\gamma_{AB} = \left(
\begin{array}{cccccccc}
 0 & 0 & -1 & 0 & 0 & 0 & 0 & 0 \\
 0 & 1/2 & 0 & 0 & 0 & 0 & 0 & 0 \\
 -1 & 0 & 0 & 0 & 0 & 0 & 0 & 0 \\
 0 & 0 & 0 & 0 & 0 & 0 & 0 & 4 \\
 0 & 0 & 0 & 0 & 0 & 0 & -1 & 0 \\
 0 & 0 & 0 & 0 & 0 & 2/3 & 0 & 0 \\
 0 & 0 & 0 & 0 & -1 & 0 & 0 & 0 \\
 0 & 0 & 0 & 4 & 0 & 0 & 0 & 0 \\
\end{array}
\right)
\end{equation}

\section{Equations of Motion}
For a general gauge field $a(t,\phi)$,

\begin{eqnarray}
a(t,\phi) &=& \left( \sum_{i=-1}^{1} \mui{i}(t,\phi)L_i + \sum_{j=-2}^{2} \nui{j}(t,\phi)W_j \right)dt \br &+& \left( \sum_{i=-1}^{1} \calL{i}(t,\phi)L_i + \sum_{j=-2}^{2} \calW{j}(t,\phi)W_j  \right)d\phi
\end{eqnarray}
equations of motion impose constraints on the chemical potentials and charges. The equation of motion is given by

\begin{equation}
F_{t\phi} = \partial_t a_{\phi} - \partial_{\phi}a_t + [a_t,a_{\phi}]
\end{equation}
which in the component form are given by

\begin{eqnarray}
\calL{1}\mui{0} - \calL{0}\mui{1} + 2\calW{1}\nui{0} &-& 2\calW{0}\nui{1} + 4\calW{-1}\nui{2} - 4\calW{2}\nui{-1} \br  && + \;\; \pp \mui{1} - \pt \calL{1} = 0
\end{eqnarray}
\begin{eqnarray}
-2\calL{-1}\mui{1} &+& 2\calL{1}\mui{-1} - 2\calW{-1}\nui{1} + 16\calW{-2}\nui{2} + 2\calW{1}\nui{-1} \br &-& 16\calW{2}\nui{-2} + \pp \nui{0} - \pt \calL{0} = 0 
\end{eqnarray}
\begin{eqnarray}
\calL{-1}\mui{0} - \calL{0}\mui{-1} + 2\calW{-1}\nui{0} &-& 4\calW{-2}\nui{1} + 2\calW{0}\nui{-1} + 4\calW{1}\nui{-2}\br && -\;\; \pp \nui{-1} + \pt \calL{-1} = 0
\end{eqnarray}
\begin{eqnarray}
2\calW{-2}\mui{0} - \calW{-1}\mui{-1} + \calL{-1}\nui{-1} + 2\calL{0}\nui{-2} - \pp \nui{-2} + \pt \calW{-2} = 0
\end{eqnarray}
\begin{eqnarray}
-\calW{-1}\mui{0} - 4\calW{-2}\mui{1} + 2\calW{0}\mui{-1} &-&  2\calL{-1}\nui{0} + \calL{0}\nui{-1} + 4\calL{1}\nui{-2} \br && +\;\; \pp \nui{-1} - \pt \calW{-1} = 0
\end{eqnarray}
\begin{eqnarray}
-3\calW{-1}\mui{1} + 3\calW{1}\mui{-1} - 3\calL{-1}\nui{1} + 3\calL{1}\nui{-1} + \pp \nui{0} - \pt \calW{0} = 0
\end{eqnarray}
\begin{eqnarray}
\calW{1}\mui{0} - 2\calW{0}\mui{1} + 4\calW{2}\mui{-1} &+& 2\calL{1}\nui{0} - \calL{0}\nui{1} - 4\calL{-1}\nui{2} \br && +\;\; \pp \nui{1} - \pt \calW{1} = 0
\end{eqnarray}
\begin{eqnarray}
2\calW{2}\mui{0} - \calW{1}\mui{1} + \calL{1}\nui{1} - 2\calL{0}\nui{2} + \pp \nui{2} - \pt\calW{2} = 0
\end{eqnarray}

\section{The Metric of \cite{nemani}}
In \cite{nemani}, the proposed gauge field solution \eqref{nemani_gauge1}, \eqref{nemani_gauge2} translates to the following metric in the Banados radial gauge $b = \exp(\rho L_0)$ that they work with:

\begin{eqnarray}
g_{\rho \rho} &=& 1,\quad g_{\rho t} = -\frac{1}{2}f^0, \quad g_{\rho \phi} = -\frac{1}{2}f^0 \br g_{tt} &=& \rme^{2\rho}(-1 + f^{-1}) + \left[ \frac{1}{4}f^{0^2} - f^1 f^{-1} + \frac{1}{3}g^{0^2} - g^{1}g^{-1} + \kappa + \tilde{\kappa} - f^{-1}\tilde{\kappa} + 4g^{-2}(g^2 + \tilde{\omega}) \right] \br &-& \rme^{-2\rho}\kappa(f^1 + \tilde{\kappa}) + 4\rme^{-4\rho}\omega(g^2 + \tilde{\omega}) \\
g_{t\phi} &=& \rme^{2\rho}f^{-1} + \left[ \frac{1}{4}f^{0^2} - f^1 f^{-1} + \frac{1}{3}g^{0^2} - g^{1}g^{-1} + \kappa - \tilde{\kappa}  + 4g^{-2}g^2 \right] - \rme^{-2\rho}f^{1}\kappa \br &+& 4\rme^{-4\rho}g^{2}\omega \br g_{\phi \phi} &=& \rme^{2\rho}(1+f^{-1}) + \left[ \frac{1}{4}f^{0^2} - f^1 (1+f^{-1}) + \frac{1}{3}g^{0^2} - g^{1}g^{-1} + \kappa + \tilde{\kappa} + f^{-1}\tilde{\kappa} + 4g^{-2}(g^2 - \tilde{\omega}) \right] \br &+& \rme^{-2\rho} \kappa(\tilde{\kappa} - f^1) + 4\rme^{-4\rho}\omega(g^2 - \tilde{\omega}) \nonumber
\end{eqnarray}
Notice that this metric does not fall under the usual Fefferman-Graham form because of the non-vanishing $g_{\rho t}$ and $g_{\rho \phi}$ components. However this metric is still allowed by the \textit{generalized Fefferman-Graham} class of metrics of \eqref{ads_metric}.

\end{document}